\documentclass[runningheads,a4paper]{llncs}
\usepackage{graphicx}
\usepackage{amsmath}
\usepackage{amssymb}
\usepackage[utf8]{inputenc}
\usepackage{url}
\usepackage{multirow}
\usepackage[linesnumbered,ruled,vlined]{algorithm2e}
\usepackage{color}

\fontfamily{ptm}
\begin{document}
\frontmatter          

\pagenumbering{arabic}
\pagestyle{headings}  

\title{Using Volunteer Computing for Mounting SAT-based Cryptographic Attacks}
%
%
\author{Alexander Semenov \and Oleg Zaikin \and Ilya Otpuschennikov}
\authorrunning{A.\,Semenov \and O.\,Zaikin \and I.\,Otpuschennikov}
\institute{Institute for System Dynamics and Control Theory SB RAS, Irkutsk, Russia
\email{biclop.rambler@yandex.ru, zaikin.icc@gmail.com, otilya@yandex.ru}}

\maketitle              

\begin{abstract}
In this paper we describe the volunteer computing project SAT@home, developed and maintained by us. This project is aimed at solving hard instances of the Boolean satisfiability problem (SAT). We believe that this project can be a useful tool for computational study of inversion problems of some cryptographic functions. In particular we describe a series of experiments performed in SAT@home on the cryptanalysis of the widely known keystream generator A5/1. In all experiments we analyzed one known burst (114 bits) of keystream produced by A5/1. Before the cryptanalysis itself there is a stage on which the partitioning of the original problem to a family of subproblems is carried out. Each of subproblems should be easy enough so that it could be solved in relatively small amount of time by volunteer's PC. We construct such partitioning using the special technique based on the Monte Carlo method and discrete optimization algorithms for special predictive functions. Besides this in the paper we describe the technique for reducing inversion problems of cryptographic functions to SAT.
\end{abstract}
\keywords{Boolean satisfiability problem, SAT, volunteer computing, BOINC, cryptanalysis, A5/1, SAT@home}
\section{Introduction}

In this paper we present our experience of the development of the volunteer computing project SAT@home aimed at solving hard instances of the Boolean satisfiability problem (SAT). In particular this project can be used to justify from the practical point of view the resistance of some cryptographic functions to attacks based on the algorithms for solving SAT. We confirm this assertion by successfully solving in SAT@home the series of cryptanalysis problems of widely known keystream generator A5/1, based on one known burst (114 bits) of keystream. 

The Boolean satisfiability problem is formulated as follows: for an arbitrary Boolean formula to determine if there exists such assignment of its variables, that makes this formula true. This problem is NP-complete, and if $P\neq NP$ there are no polynomial algorithms that could solve it. On the other hand, the class of practically important problems that can be effectively reduced to SAT is very wide \cite{DBLP:series/faia/2009-185}. Because of this fact it is important to develop the algorithms that show good effectiveness in application to various cases of SAT.

In practice it is usually convenient to consider satisfiability problem for a formula presented in the Conjunctive Normal Form (CNF). Using the Tseitin transformations \cite{Tseitin83} it is possible to effectively (in polynomial time) reduce the SAT for an arbitrary Boolean formula to the SAT for some CNF.

In the last 10 years the effectiveness of SAT solving algorithms has significantly increased. Today SAT approach is used to solve various problems from the areas of symbolic verification, bioinformatics, discrete optimization, combinatorial design, cryptanalysis, etc. There are several basic classes of algorithms used to construct modern SAT solvers. All the results presented in this paper were obtained using the so called Conflict Driven Clause Learning (CDCL) solvers, the description of main architectural principles of which can be found, for example, in \cite{DBLP:series/faia/SilvaLM09}.

To solve SAT instances encoding the problems of cryptanalysis of ciphering systems used in the real world it is insufficient to employ only the resources of a single PC. In this context the development of SAT solving methods, designed for parallel and distributed computing environments, is highly relevant. There are several general concepts of solving SAT in parallel. From our point of view for solving SAT instances, encoding cryptographic attacks, the volunteer computing concept looks promising, because this kind of distributed computing makes it possible to solve one particular computing problem for a prolonged time with relatively high performance (on average).

It should be noted, that when solving a particular problem in a volunteer computing project there apply some restrictions. First, the problem should be divided into subproblems in such a way that each subproblem can be solved by a volunteer's PC in relatively small amount of time (several hours at most). Also it is worth mentioning that volunteers PCs do not interconnect directly with each other. From these restrictions it is natural to draw the conclusion that to parallelize the original problem it is best to use the partitioning approach. The partitioning method must make it possible to estimate the time required to solve all produced subproblems. Below we describe such method and apply it to the parallelization of cryptanalysis problems for the widely known keystream generator A5/1. Then we process the partitionings obtained in parallel using the volunteer computing project SAT@home.

Let us give a brief outline of the paper. In the second section we present the general technique used to reduce cryptographic problems to SAT and describe the \textsc{Transalg} program system implementing this technique. In the third section we give brief description of main SAT solving algorithms that we applied to cryptographic problems. In the fourth section we propose an approach to the parallelization of SAT instances that encode cryptanalysis problems. Thus, we describe the general technique of constructing SAT partitionings that can be processed in distributed computing environments. In the fifth section we present the results of the computational experiments on the search for SAT partitionings, with the use of the algorithms described in the previous section. In the sixth section we describe the volunteer computing project SAT@home developed by us specifically for the purpose of solving hard SAT instances, and present the results of computational experiments held in SAT@home on the cryptanalysis of the A5/1 keystream generator.

\section{Reducing Inversion Problems of Cryptographic Functions to SAT}

Hereinafter by $\{0,1\}^l$ we denote the set formed by all possible binary words of length $l$. By $f:\{0,1\}^n\rightarrow\{0,1\}^m$ we denote a discrete function that transforms binary words of length $n$ to binary words of length $m$. For each function considered below we assume that this function is defined everywhere on $\{0,1\}^n$ and is specified by some algorithm $A$. By $Range\;f$, $Range\;f\subseteq\{0,1\}^m$ we denote the set of all output values of the function $f$. Let us consider the following problem: for a given $y\in Range\;f$ and the known algorithm $A$ we need to find such $x\in\{0,1\}^n$, that $f(x)=y$. We will refer to this general problem as to the problem of inversion of function $f$. Many cryptanalysis problems can be considered in its context. For example, let $f$ be a function specified by the algorithm of some keystream generator. Then $x\in\{0,1\}^n$ is the secret key and $y\in\{0,1\}^m:f(x)=y$ is the keystream produced by the generator considered. In the vast majority of keystream ciphers the ciphertext is formed by adding keystream and some secret message $M$ in arithmetics modulo 2. Sometimes, it is possible (usually when the protocol is inconsistent) that some message $M$ becomes known. It may represent some technical data and thus not even contain any valuable information (as in case of the ciphering protocol used in the GSM standard). However, if we know $M$ and the corresponding ciphertext then we can find $y$ and consider the problem of search for the secret key $x:f(x)=y$. Knowing $x$ makes it possible to read any message ciphered using this secret key. The problem of determining $x$ based on a given $y$ and known algorithm specifying function $f$ is usually called a known plaintext attack. Hereinafter we will study this particular cryptographic problem.

\subsection{Theoretical Basis of SAT Encoding of Discrete Functions Inversion Problems}

Let us describe the general scheme of encoding the problem of inversion of $f$ to SAT. The main idea that underlies such reductions is based on the Cook theorem \cite{DBLP:conf/stoc/Cook71}. It consists in forming a correspondence between the computation of function $f$ by algorithm $A$ and some system of Boolean equations. The complexity of the procedure that constructs the system of Boolean equations is bounded above by a polynomial of the complexity of algorithm $A$.

In more detail the transition from the problem of inversion of $f$ to the CNF encoding looks as follows. By $S(f)$ we denote the Boolean Circuit over the basis $\{\wedge,\neg\}$ that implements function $f$. In the circuit $S(f)$ there are $n$ inputs and $m$ outputs. Let us link Boolean variables $x_1,\ldots,x_n$, forming the set $X$, with inputs of the circuit, and Boolean variables $y_1,\ldots,y_m$, forming the set $Y$, with its outputs. We also mark each logical gate $G$ by an auxiliary variable $v(G)$ such that $v(G)\notin X$. The set of all auxiliary variables we denote by $V$. Note that $Y\subseteq V$. Let $v(G)$ be an arbitrary variable from $V$ and $G$ be a corresponding gate. If $G$ is a NOT-gate and $u\in X\bigcup V$ is the variable linked with the input of $G$ then we encode the gate $G$ with the Boolean formula $v(G)\equiv \neg u$ (hereinafter by $\equiv$ we mean logical equivalence). If $G$ is an AND-gate and $u,w\in X\bigcup V$ are variables linked with the gate inputs then we encode $G$ with the formula $v(G)\equiv u\wedge w$. CNF-representations of Boolean functions specified by formulas $v(G)\equiv \neg u$ and $v(G)\equiv u\wedge v$ look as follows (respectively):
\begin{equation}
\label{f1}
\begin{array}{l}
\left(v(G)\vee u\right)\wedge \left(\neg v(G)\vee \neg u\right)\\
\left(v(G)\vee \neg u\vee \neg w\right)\wedge \left(\neg v(G)\vee u \right)\wedge \left(\neg v(G)\vee w\right)
\end{array}
\end{equation}
Thus with an arbitrary gate $G$ of circuit $S(f)$ we associate a CNF $C(G)$ of the kind \eqref{f1}. We will say that the CNF $C(G)$ encodes the gate $G$, and the CNF 
\begin{equation*}
C(f)=\bigwedge\limits_{G\in S(f)}C(G)
\end{equation*}
encodes the circuit $S(f)$. The described technique of constructing a CNF for a circuit $S(f)$ is known as Tseitin transformations \cite{Tseitin83}. 

Now let $\left(\beta_1,\ldots,\beta_m\right)\in\{0,1\}^m$ be an arbitrary assignment of variables $y_1,\ldots,y_m$. Consider the CNF
\begin{equation}
\label{f2}
C(f)\wedge y_1^{\beta_1}\wedge\ldots\wedge y_m^{\beta_m}
\end{equation}
where 
\begin{equation*}
y^{\beta}=\left\{
\begin{array}{cc}
\neg y,& if \;\beta=0\\
y,& if\;\beta=1
\end{array}
\right.
\end{equation*}
From the properties of the Tseitin transformations it follows that if $\left(\beta_1,\ldots,\beta_m\right)\in Range\;f$ then the CNF \eqref{f2} is satisfiable and from any of its satisfying assignments one can effectively extract such assignment $x_1=\alpha_1,\ldots,x_n=\alpha_n$, $\alpha_i\in\{0,1\}$, $i\in \{1,\ldots,n\}$, that $f\left(\alpha_1,\ldots,\alpha_n\right)=\left(\beta_1,\ldots,\beta_m\right)$.

Thus, using the technique described above we reduce the problem of inversion of function $f$ in an arbitrary point $y\in Range f$ to SAT for the CNF \eqref{f2}. It makes it possible to apply the SAT solvers to the latter problem.

\subsection{Encoding Inversion Problems of Cryptographic Functions to SAT Using the \textbf{Transalg} System}

In this section we will briefly describe the \textsc{Transalg} program system \cite{DBLP:journals/corr/OtpuschennikovSK14} that was designed specifically to reduce inversion problems of cryptographic functions to SAT. The original function $f$ should first be represented in the form of the special TA-program $A$. TA-programs have C-like syntax. The translation of an arbitrary TA-program $A$ essentially involves the symbolic execution of $A$ \cite{King:1976:SEP:360248.360252}, and results in the propositional encoding of $A$ in the form of CNF. In other words the result of compilation of the program $A$ is not a machine code but a set of Boolean formulas constructed according to the Tseitin transformations. In more detail, the result of the compilation of the TA-program is the CNF $C(f)$ encoding the circuit $S(f)$. By fixing the values of output variables $\left(\beta_1,\ldots,\beta_m\right)$ (essentially transitioning to CNF \eqref{f2}) we obtain a SAT instance that encodes the inversion problem of function $f$ in the point $y=\left(\beta_1,\ldots,\beta_m\right)$.

\begin{figure}[ht]
	\centering
		\includegraphics[width=7.5cm]{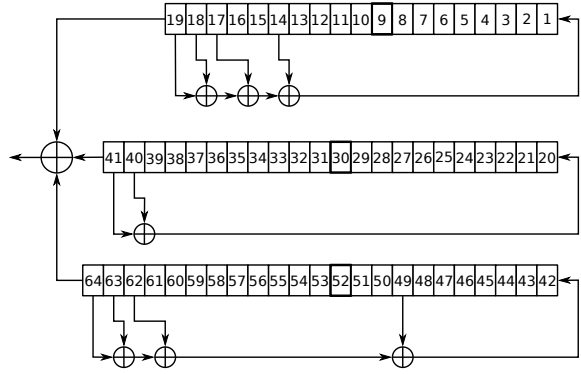}
	\caption{The A5/1 keystream generator}
	\label{a5_1_scheme}
\end{figure}

Below the main object of our study is the A5/1 keystream generator that has been used to encrypt data within the GSM standard for more than 15 years.
On Figure \ref{a5_1_scheme} we demonstrate the scheme of the A5/1 generator from the paper \cite{DBLP:conf/fse/BiryukovSW00}. 

The generator contains three linear feedback shift registers (LFSRs) specified by the following feedback polynomials:
\begin{equation*}
\begin{array}{ll}
LFSR1 & X^{19}+X^{18}+X^{17}+X^{14}+1;\\
LFSR2 & X^{22}+X^{21}+1;\\
LFSR3 & X^{23}+X^{22}+X^{21}+X^8+1.
\end{array}
\end{equation*}

The registers in A5/1 are shifted asynchronously. Assume that at the initial time moment $t=0$ the registers contain the secret key. By $b_i^t$, $i\in\{1,2,3\}$ denote the value of the clocking bit of the $i$-th register at time moment $t$. The clocking bits are contained in cells $9$, $30$ and $52$. With each clocking bit the special function
\begin{equation*}
\tau_i(t)=\left\{
\begin{array}{l}
1, if\;majority\;\left(b_1(t),b_2(t),b_3(t)\right)=b_i(t)\\
0, if\;majority\;\left(b_1(t),b_2(t),b_3(t)\right)\neq b_i(t)
\end{array}
\right.
\end{equation*}
is associated. Here $majority\left(A,B,C\right)=(A\wedge B)\vee (A\wedge C)\vee (B\wedge C)$. The register number $i\in\{1,2,3\}$ is shifted at time moment $t\in\{1,2,\ldots\}$ only if $\tau_i(t)=1$. Otherwise the register retains the condition in which it was at the previous time moment. At each time moment the generator outputs one bit of keystream that is produced as a sum modulo 2 of bits contained in cells $19$, $41$ and $64$. The length of the secret key in the A5/1 keystream generator is $64$ bits.

\begin{figure}[ht]
	\centering
		\includegraphics[width=10cm]{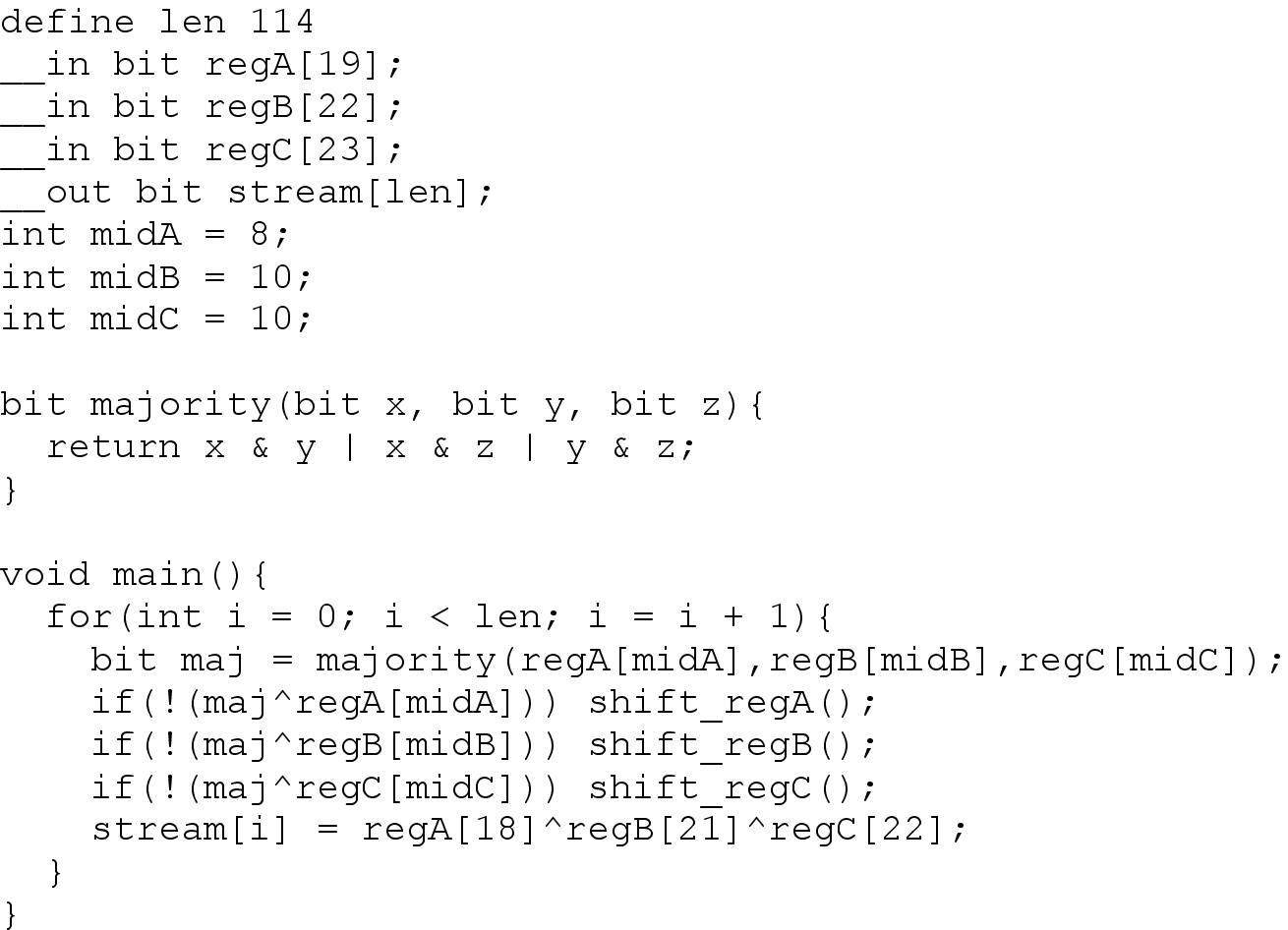}
	\caption{TA-program specifying function $f_{A5/1}$.}
	\label{a5_1_ta_program}
\end{figure}

On Figure \ref{a5_1_ta_program} we present the TA-program specifying the discrete function
\begin{equation}
\label{f3}
f_{A5/1}:\{0,1\}^{64}\rightarrow\{0,1\}^{114}.
\end{equation}
This function generates 114 bits of keystream from the secret key of length 64 bits according to the A5/1 algorithm. The length of keystream fragment is chosen to be equal to 114 bits because in the GSM standard the information is encrypted by blocks of 114 bits. Such blocks are called bursts. 

The result of the compilation of the program from the Figure \ref{a5_1_ta_program} is the CNF with 7816 variables and 35568 clauses (size of the file in DIMACS format is 730 kilobytes).

\section{Algorithms for Solving SAT Instances that Encode Inversion Problems of Cryptographic Functions}

Main classes of the SAT solving algorithms were extensively described in \cite{DBLP:series/faia/2009-185}. After many computational experiments we made the conclusion that the algorithms based on the concept of Conflict Driven Clause Learning (CDCL) \cite{DBLP:series/faia/SilvaLM09} perform best on cryptographic problems. All the modern CDCL solvers that show good effectiveness on wide classes of industrial tests are based on the DPLL algorithm \cite{Davis:1962:MPT:368273.368557}. However, in its pure form the DPLL algorithm has relatively low performance. The main idea of the CDCL concept is to complement the DPLL with an ability to use memory to store the information about traversed parts of the search tree. This information is stored in the form of restrictive constraints which are called conflict clauses. The solver GRASP \cite{DBLP:journals/tc/Marques-SilvaS99} was the first full-fledged CDCL solver. In the subsequent papers \cite{DBLP:conf/iccad/ZhangMMM01,DBLP:conf/dac/MoskewiczMZZM01} several important algorithmical constructions were described that made it possible to improve the CDCL concept. It led to the creation of SAT solvers widely applied nowadays to combinatorial problems from various areas. 

Active development of parallel SAT solving algorithms started quite recently. First parallel SAT solvers competition was held in 2008. There are two main approaches to parallel SAT solving: the partitioning approach and the portfolio approach. In the partitioning approach the search space is divided into separate subspaces that are processed in parallel. In the portfolio approach all computing processes work with the same original search space, but they use different initial parameters and search heuristics, and also can share the collected data with each other.

The SAT instances encoding the inversion problems of cryptographic functions used in the real world usually are very hard, and to solve them it is insufficient to use the resources of a usual PC. The graphical processors (GPU) in the context of solving SAT nowadays do not yield any advantages in comparison with traditional CPUs. That is explained by the fact that modern GPUs have relatively high memory latencies, which significantly decrease the performance of CDCL algorithms. 

For solving hard SAT instances one can use computing clusters with nodes based on CPUs. The portfolio approach makes it impossible to effectively utilize a large number of cluster nodes because intensive conflict clauses exchange greatly taxes the interconnection environment. The advantage of the partitioning approach is that it doesn't limit the amount of computating nodes of the cluster that can be used. The low intensity of data exchange between computing processes in this case makes it possible to employ not only computing clusters but also distributed computing systems. The use of such systems (including volunteer computing projects) suits better for solving hard SAT instances because of the low computing costs (in comparison to that of clusters). To organize the computations in the distributed system the original problem should be divided into a family of significantly less complex subproblems in such a way that each subproblem can be solved by a computing node (such as, for example, a usual PC) in relatively small amount of time. In the next section we will describe the SAT partitioning technique that we used to organize the computations in grid systems.

\section{Partitioning Procedures for SAT Instances Encoding Inversion Problems of Cryptographic Functions}

So as we explained above we use the partitioning approach to parallelize the SAT solving process. There exist many various ways of constructing SAT partitionings \cite{Hyvarinen11}. Moreover, even following one way it is possible to make several different partitionings for the same SAT instance. The main problem that arises in this context is to justify why one partitioning is better than another. The basic partitioning effectiveness criterion is the time required to process it in a distributed computing environment. Unfortunately, not for all types of SAT partitionings it is possible to evaluate their effectiveness. In this section we will describe one type of SAT partitioning for which there exists quite natural procedure for constructing estimations of time required to process it. We used this type of partitioning to solve inversion problems of some cryptographic functions in parallel.

Assume that $C$ is the CNF encoding the inversion problem of some discrete function. By $X$ denote the set of Boolean variables that appear in CNF $C$. By $\tilde{X}$ denote an arbitrary subset of $X$. Let $\alpha$ be an arbitrary set of values of variables from $\tilde{X}$. By $C\left[\tilde{X}/\alpha\right]$ we denote the CNF produced from $C$ by setting values from $\alpha$ to corresponding variables. Let us consider the set $\Delta\left(C,\tilde{X}\right)=\left\{C\left[\tilde{X}/\alpha\right]\right\}_{\alpha\in\{0,1\}^{|\tilde{X}|}}$. It is easy to see that the CNF $C$ is satisfiable if and only if at least one CNF from the set $\Delta\left(C,\tilde{X}\right)$ is satisfiable. Therefore, the set $\Delta\left(C,\tilde{X}\right)$ is a partitioning of the SAT instance $C$ \cite{Hyvarinen11}. We call the set $\tilde{X}$ that produces the partitioning $\Delta\left(C,\tilde{X}\right)$ a \textit{decomposition set}.

Now let us describe the approach to constructing time estimations for the processing of SAT partitionings that was proposed in \cite{DBLP:journals/corr/SemenovZ13}. First let us fix some particular SAT solving algorithm $A$. Hereinafter we assume that $A$ is a complete algorithm, i.e. it has finite runtime for an arbitrary input. Now consider an arbitrary CNF $C$ over the set $X$ of Boolean variables. Let $\tilde{X}$ be an arbitrary decomposition set. Consider the set $\{0,1\}^{|\tilde{X}|}$. Let us define the uniform distribution on this set. With a randomly selected vector $\alpha\in\{0,1\}^{|\tilde{X}|}$ we associate the runtime of algorithm $A$ given the CNF $C\left[\tilde{X}/\alpha\right]$ as an input. This time can be considered as a value of some random variable which we denote below as $\xi_A\left(C,\tilde{X}\right)$. Since $A$ is a complete algorithm, then the random variable $\xi_A\left(C,\tilde{X}\right)$ takes only finite values with some probabilities. Thus, the expected value of $\xi_A\left(C,\tilde{X}\right)$ is finite. By $t_A\left(C,\tilde{X}\right)$ denote the time required to process the SAT partitioning $\Delta\left(C,\tilde{X}\right)$ sequentially by algorithm $A$. In \cite{DBLP:journals/corr/SemenovZ13} it was shown that under listed conditions the following equality holds:
\begin{equation}
\label{f4}
t_A\left(C,\tilde{X}\right)=2^{|\tilde{X}|}\cdot\mathrm{E}\left[\xi_A\left(C,\tilde{X}\right)\right]
\end{equation}

To estimate the expected value $\mathrm{E}\left[\xi_A\left(C,\tilde{X}\right)\right]$ it is possible to use the computational scheme of the Monte Carlo method in its classical form \cite{Metropolis49}. Namely, for an arbitrary decomposition set $\tilde{X}$ we randomly select vectors $\alpha_1,\ldots,\alpha_N$ from $\{0,1\}^{|\tilde{X}|}$. Then for each $\alpha_j$, $j\in\{1,\ldots,N\}$ we measure the time required by algorithm $A$ to solve the SAT for CNF $C\left[\tilde{X}/\alpha_j\right]$. We denote the time obtained by $\xi^j$. After this we compute the value $\frac{1}{N}\cdot\sum\limits_{j=1}^{N}\xi^j$. In accordance with the theoretical basis of the Monte Carlo method \cite{Metropolis49,Kalos:109491}, when $N$ is large enough, the value $\frac{1}{N}\cdot\sum\limits_{j=1}^{N}\xi^j$ becomes a good estimation of the expected value $\mathrm{E}\left[\xi_A\left(C,\tilde{X}\right)\right]$. Therefore it follows from \eqref{f4} that the value
\begin{equation}
\label{f5}
F_A\left(C,\tilde{X}\right)=\frac{2^{|\tilde{X}|}}{N}\cdot\sum\limits_{j=1}^{N}\xi^j
\end{equation}
is the estimation of time required by algorithm $A$ to sequentially process the whole SAT partitioning $\Delta\left(C,\tilde{X}\right)$ and this estimation is the more accurate the larger $N$ is. Below we will refer to the function defined according to \eqref{f5} as to the \textit{predictive function}.

As we already noted above, different partitionings of the same SAT problem can have different values of $t_A\left(C,\tilde{X}\right)$. In practice it is important to be able to find partitionings that can be processed in affordable time. Below we will describe a scheme of automatic search for good partitionings that is based on the procedure minimizing the predictive function value in a special search space.

So we consider the satisfiability problem for some CNF $C$. Let $X=\left\{x_1,\ldots,x_M\right\}$ be a set of all Boolean variables in this CNF and $\tilde{X}\subseteq X$ be an arbitrary decomposition set. The set $\tilde{X}$ can be represented by the binary vector $\chi=\left(\chi_1,\ldots,\chi_M\right)$. Here 
\begin{equation*}
\chi_i=\left\{
\begin{array}{l}
1, if\;x_i\in\tilde{X}\\
0, if\;x_i\notin\tilde{X}
\end{array}
\right.
,i\in\{1,\ldots,M\}
\end{equation*}
With an arbitrary vector $\chi\in\{0,1\}^M$ we associate the value of function $F(\chi)$ computed in the following manner. For vector $\chi$ we construct the corresponding set $\tilde{X}$ (it is formed by variables from $X$ that correspond to $1$ positions in $\chi$). Then we generate a random sample $\{\alpha_1,\ldots,\alpha_N\}$, $\alpha_j\in\{0,1\}^{|\tilde{X}|}$ and solve SAT for CNFs $C\left[\tilde{X}/\alpha_j\right]$. For each of these SAT instances we measure $\xi^j$ --- the runtime of algorithm $A$ on the input $C\left[\tilde{X}/\alpha_j\right]$. After this we calculate the value of $F_A\left(C,\tilde{X}\right)$ according to \eqref{f5}. As a result we have the value of $F(\chi)$ in the considered point of the search space.

Now we will solve the problem $F(\chi)\rightarrow min$ over the set $\{0,1\}^M$. Of course, the problem of search for the exact minimum of function $F(\chi)$ is extraordinarily complex. Therefore our main goal is to find in affordable time the points in $\{0,1\}^M$ with relatively good values of function $F(\cdot)$. For this purpose we propose two minimization algorithms. The first employs the simulated annealing algorithm, while the second is based on the tabu search concept. We will compare the effectiveness of these algorithms on the example of the problem of search for the good SAT partitioning for the SAT instances encoding the cryptanalysis of the A5/1 keystream generator (i.e. the inversion problem of \eqref{f3}). 

First we need to introduce the notation. By $\Re$ we denote the search space, for example, $\Re=\{0,1\}^M$, however, as we will see later, for the problems considered one can use the search spaces of much less power. The minimization of function $F(\cdot)$ is considered as an iterative process of transitioning between the points of the search space:
\begin{equation*}
\chi^0\rightarrow\chi^1\rightarrow\ldots\rightarrow\chi^i\rightarrow\ldots\rightarrow\chi^{\ast}.
\end{equation*}
By $N_{\rho}\left(\chi\right)$ we denote the neighborhood of point $\chi$ of radius $\rho$ in the search space $\Re$. The point from which the search starts we denote as $\chi_{start}$. The current Best Known Value of $F(\cdot)$ is denoted by $F_{best}$. The point in which the $F_{best}$ was achieved we denote as $\chi_{best}$. By $\chi_{center}$ we denote the point the neighborhood of which is processed at the current moment. We call the point, in which we computed the value $F(\cdot)$ a \textit{checked point}. The neighborhood $N_{\rho}\left(\chi\right)$ in which all the points are checked is called \textit{checked neighborhood}. Otherwise the neighborhood is called \textit{unchecked}. 

According to the scheme of the simulated annealing \cite{Kirkpatrick83optimizationby}, the transition from $\chi^i$ to $\chi^{i+1}$ is performed in two stages. First we choose a point $\tilde{\chi}^i$ from $N_{\rho}\left(\chi^i\right)$. The point $\tilde{\chi}^i$ becomes the point $\chi^{i+1}$ with the probability denoted as $\mathrm{Pr}\left\{\tilde{\chi}^i\rightarrow\chi^{i+1}|\chi^i\right\}$. This probability is defined in the following way:
\begin{equation*}
\mathrm{Pr}\left\{\tilde{\chi}^i\rightarrow\chi^{i+1}|\chi^i\right\}=\left\{
\begin{array}{cc}
1,& if\;F\left(\tilde{\chi}^i\right)<F\left(\chi^i\right)\\
\exp\left( -\frac{F\left(\tilde{\chi}^i\right)-F\left(\chi^i\right)}{T_i}\right),& if\;F\left(\tilde{\chi}^i\right)\geq F\left(\chi^i\right)
\end{array}
\right.
\end{equation*}
In the pseudocode of the algorithm demonstrated below the function that tests if the point $\tilde{\chi}^i$ becomes $\chi^{i+1}$, is called \texttt{PointAccepted} (this function returns the value of \texttt{true} if the transition occurs and \texttt{false} otherwise). The change of parameter $T_i$ corresponds to decreasing the ``temperature of the environment'' \cite{Kirkpatrick83optimizationby} (in the pseudocode by \texttt{decreaseTemperature()} we denote the function which implements this procedure). Usually it is assumed that $T_i=Q\cdot T_{i-1}$, $i\geq 1$, where $Q\in(0,1)$. The process starts at some initial value $T_0$ and continues until the temperature drops below some threshold value $T_{inf}$ (in the pseudocode the function that checks this condition is called \texttt{temperatureLimitReached()}).

\begin{algorithm}[htb]
 \DontPrintSemicolon
 \SetKwData{false}{false}
 \SetKwData{true}{true}
 \SetKwData{bestValueUpdated}{bestValueUpdated}
 \SetKwFunction{PointAccepted}{PointAccepted}
 \SetKwFunction{timeExceeded}{timeExceeded}
 \SetKwFunction{temperatureLimitReached}{temperatureLimitReached}
 \SetKwFunction{decreaseTemperature}{decreaseTemperature}
 \caption{Simulated annealing algorithm for minimization of the predictive function}
 \KwIn{CNF $C$, initial point $\chi_{start}$}
 \KwOut{Pair $\langle \chi_{best}, F_{best} \rangle$, where $F_{best}$ is a prediction for $C$, $\chi_{best}$ is a corresponding decomposition set}
	$\langle \chi_{center}, F_{best} \rangle \gets \langle \chi_{start}, F(\chi_{start}) \rangle$\;
	\Repeat{\timeExceeded{} or \temperatureLimitReached{}} {
		\bestValueUpdated $\gets$ \false\;
		$\rho = 1$\;
		\Repeat(\tcp*[f]{check neighborhood}){\bestValueUpdated}{
			$\chi \gets$ any unchecked  point from $N_{\rho}(\chi_{center})$ \;
			compute $F(\chi)$\;
			mark $\chi$ as checked point in $N_{\rho}(\chi_{center})$\;
			\If{\PointAccepted{$\chi$}}{
				$\langle \chi_{best}, F_{best} \rangle \gets \langle \chi, F(\chi) \rangle$\;
				$\chi_{center} \gets \chi_{best}$\;
				\bestValueUpdated $\gets$ \true\;
			}
			\If{($N_{\rho}(\chi_{center})$ is checked) and (not \bestValueUpdated)}{
				$\rho = \rho + 1$\;
			}
			\decreaseTemperature{}\;
		}
	}
\Return{$\langle \chi_{best}, F_{best} \rangle$}\;
\end{algorithm}

In the other approach to the minimization of $F(\cdot)$ we employed the tabu search scheme \cite{Glover:1997:TS:549765}. According to this approach the points from the search space, in which we already calculated the values of function $F(\cdot)$ are stored in special tabu lists. When we try to improve the current Best Known Value of $F(\cdot)$ in the neighborhood of some point $\chi_{center}$ then for an arbitrary point $\chi$ from the neighborhood we first check if we haven’t computed $F(\chi)$ earlier. If we haven’t and, therefore, the point $\chi$ is not contained in tabu lists, then we compute $F(\chi)$. This strategy is justified in the case of the minimization of predictive function $F(\cdot)$ because the computing of values of the function in some points of the search space is very expensive. The use of tabu lists makes it possible to significantly increase the number of points of the search space processed per time unit.

Let us describe the Tabu Search algorithm (TS-algorithm) for minimization $F(\cdot)$ in more detail. To store the information about points, in which we already computed the value of $F(\cdot)$ we use two tabu lists $L_1$ and $L_2$. The $L_1$ list contains only points with checked neighborhoods. The $L_2$ list contains checked points with unchecked neighborhoods. Below we present the pseudocode of the TS-algorithm for $F(\cdot)$ minimization. 

\begin{algorithm}
 \DontPrintSemicolon
 \SetKwData{false}{false}
 \SetKwData{true}{true}
 \SetKwData{bestValueUpdated}{bestValueUpdated}
 \SetKwFunction{markPointInTabuLists}{markPointInTabuLists}
 \SetKwFunction{getPoint}{getPoint}
 \SetKwFunction{timeExceeded}{timeExceeded}
 \caption{Tabu search altorithm for minimization of the predictive function}
 \KwIn{CNF $C$, initial point $\chi_{start}$}
 \KwOut{Pair $\langle \chi_{best}, F_{best} \rangle$, where $F_{best}$ is a prediction for $C$, $\chi_{best}$ is a corresponding decomposition set}
	$\langle \chi_{center}, F_{best} \rangle \gets \langle \chi_{start}, F(\chi_{start}) \rangle$\;
	$\langle L_1, L_2 \rangle \gets \langle \emptyset, \chi_{start} \rangle$ \tcp*[r]{initialize tabu lists}
	\Repeat{\timeExceeded{} or $L_2 = \emptyset$} {
		\bestValueUpdated $\gets$ \false\;
		\Repeat(\tcp*[f]{check neighborhood}){$N_{\rho}(\chi_{center})$ is checked}{
			$\chi \gets$ any unchecked point from $N_{\rho}(\chi_{center})$\;
			compute $F(\chi)$\;
			\markPointInTabuLists{$\chi, L_1, L_2$} \tcp*[r]{update tabu lists}
			\If{$F(\chi) < F_{best}$}{
				$\langle \chi_{best}, F_{best} \rangle \gets \langle \chi, F(\chi) \rangle$\;
				\bestValueUpdated $\gets$ \true\;
			}
		}
		\lIf{\bestValueUpdated} {
			$\chi_{center} \gets \chi_{best}$\;
		}
		\lElse{
			$\chi_{center} \gets$ \getPoint{$L_2$}\;
		}
	}
\Return{$\langle \chi_{best}, F_{best} \rangle$}\;
\end{algorithm}

In this algorithm the function \texttt{markPointInTabuLists$\left(\chi,L_1,L_2\right)$} adds the point $\chi$ to $L_2$ and then marks $\chi$ as checked in all neighborhoods of points from $L_2$ that contain $\chi$. If as a result the neighborhood of some point $\chi'$ becomes checked, the point $\chi'$ is removed from $L_2$ and is added to $L_1$. If we have processed all the points in the neighborhood of $\chi_{center}$ but could not improve the $F_{best}$ then as the new point $\chi_{center}$ we choose some point from $L_2$. It is done via the function \texttt{getPoint($L_2$)}. To choose the new point in this case one can use various heuristics. At the moment the TS-algorithm employs the following heuristics: it chooses the point for which the total conflict activity of Boolean variables, contained in the corresponding decomposition set, is the largest.

As we already mentioned above, taking into account the features of the considered SAT problems makes it possible to significantly decrease the size of the search space. For example, knowing the so called Backdoor Sets \cite{DBLP:conf/ijcai/WilliamsGS03} can help in that matter. In case of the SAT instances encoding the inversion problems of functions of the kind $f:\{0,1\}^n\rightarrow\{0,1\}^m$ the set $\tilde{X}_{in}$ formed by the variables encoding the inputs of the Boolean circuit $S(f)$ is the so called Strong Unit Propagation Backdoor Set \cite{DBLP:journals/constraints/JarvisaloJ09}. It means that if we use $\tilde{X}_{in}$ as a decomposition set, then the CDCL solver will solve SAT for any CNF of the kind $C\left[\tilde{X}_{in}/\alpha\right]$, $\alpha\in\{0,1\}^{|\tilde{X}_{in}|}$ on the preprocessing stage, i.e. very fast. Therefore the point $\chi_{in}$ that corresponds to the set $\tilde{X}_{in}$ can be used as an initial point in the predictive function minimization procedure. Moreover, it is possible to reduce the search space to the set $2^{\tilde{X}_{in}}$. In all our computational experiments we followed this path. 

\section{Computing Cluster Implementation of the Monte Carlo Algorithms for Constructing SAT Partitionings}

The minimization algorithms suggested above were implemented as a parallel MPI-program \textsc{\textsc{PDSAT}} \cite{DBLP:journals/corr/SemenovZ13}. In \textsc{\textsc{PDSAT}} there is one leader process, all the other are follower processes (each process corresponds to 1 CPU core).The leader process in the \textsc{PDSAT} program selects point $\chi$ of the search space and creates random sample $\{\alpha_1,\ldots,\alpha_N\}$. Follower processes solve corresponding SAT instances of the kind $C\left[\tilde{X}/\alpha_j\right]$, $j\in\{1,\ldots,N\}$ (see section 4). The value of predictive function is always computed assuming that all the subproblems from the partitioning $\Delta\left(C,\tilde{X}\right)$ will be processed by 1 CPU core. 

We considered the inversion problem of the function \eqref{f3}: given the 114 bits of keystream we needed to find the secret key of length 64 bits, which produces this keystream (in accordance with the A5/1 algorithm). The \textsc{\textsc{PDSAT}} program was used to find partitionings with good time estimations for CNFs encoding this problem. The computational experiments were performed on the computing cluster Blackford ISDCT SB RAS\footnote{http://hpc.icc.ru/} --– in each experiment \textsc{\textsc{PDSAT}} was launched for 1 day using 16 Intel Xeon 5345 EM64T CPUs (64 cores in total). We used random samples of size $N = 10^4$.

\begin{figure}[!ht]
	\centering
		\includegraphics[width=7.5cm]{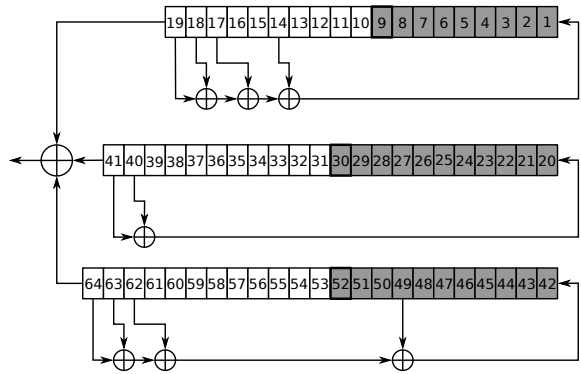}
	\caption{Decomposition set S1}
	\label{a5_1_set_S1}
\end{figure}

\begin{figure}[!ht]
	\centering
		\includegraphics[width=7.5cm]{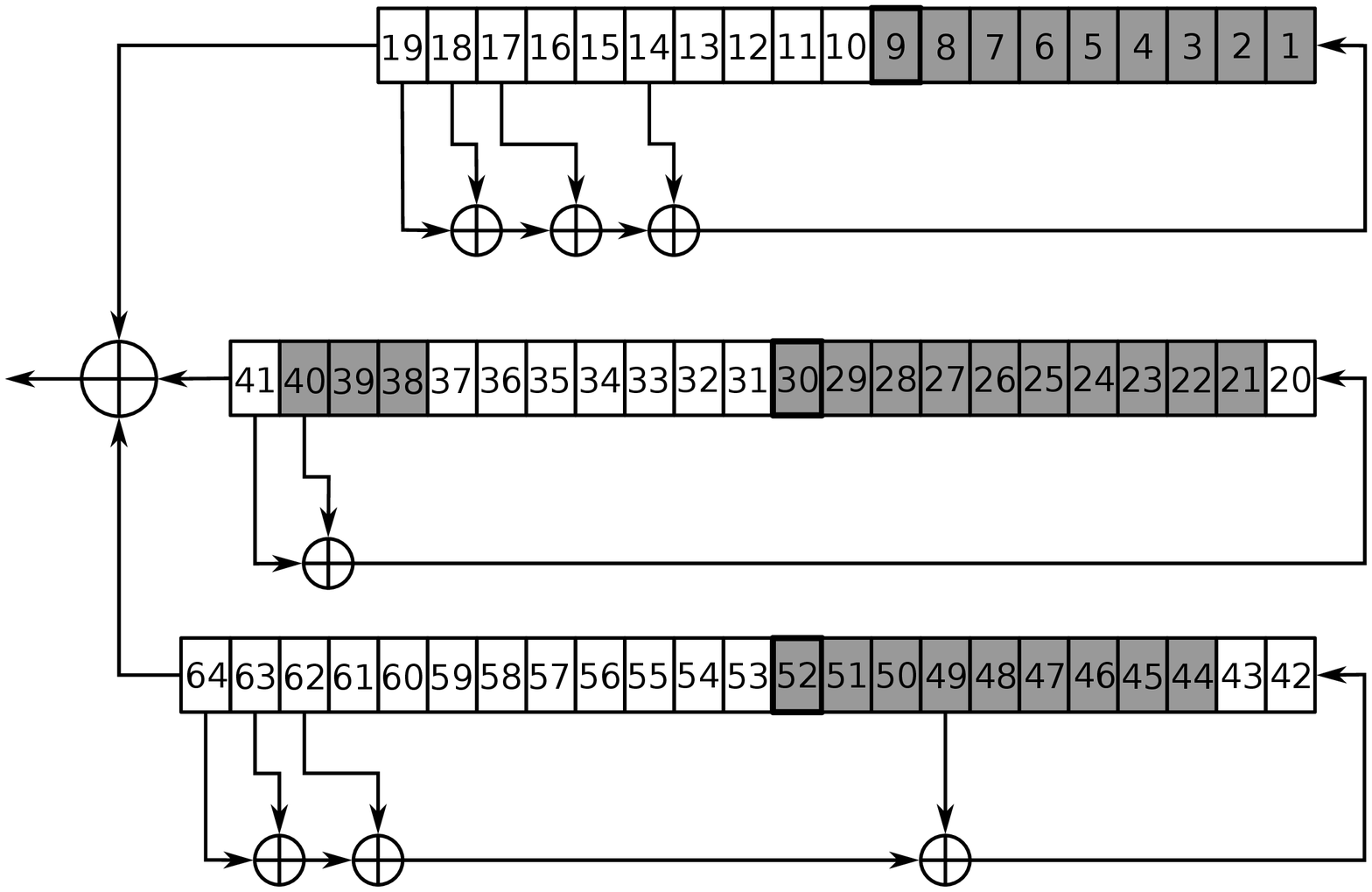}
	\caption{Decomposition set S2}
	\label{a5_1_set_S2}
\end{figure}

\begin{figure}[!ht]
	\centering
		\includegraphics[width=7.5cm]{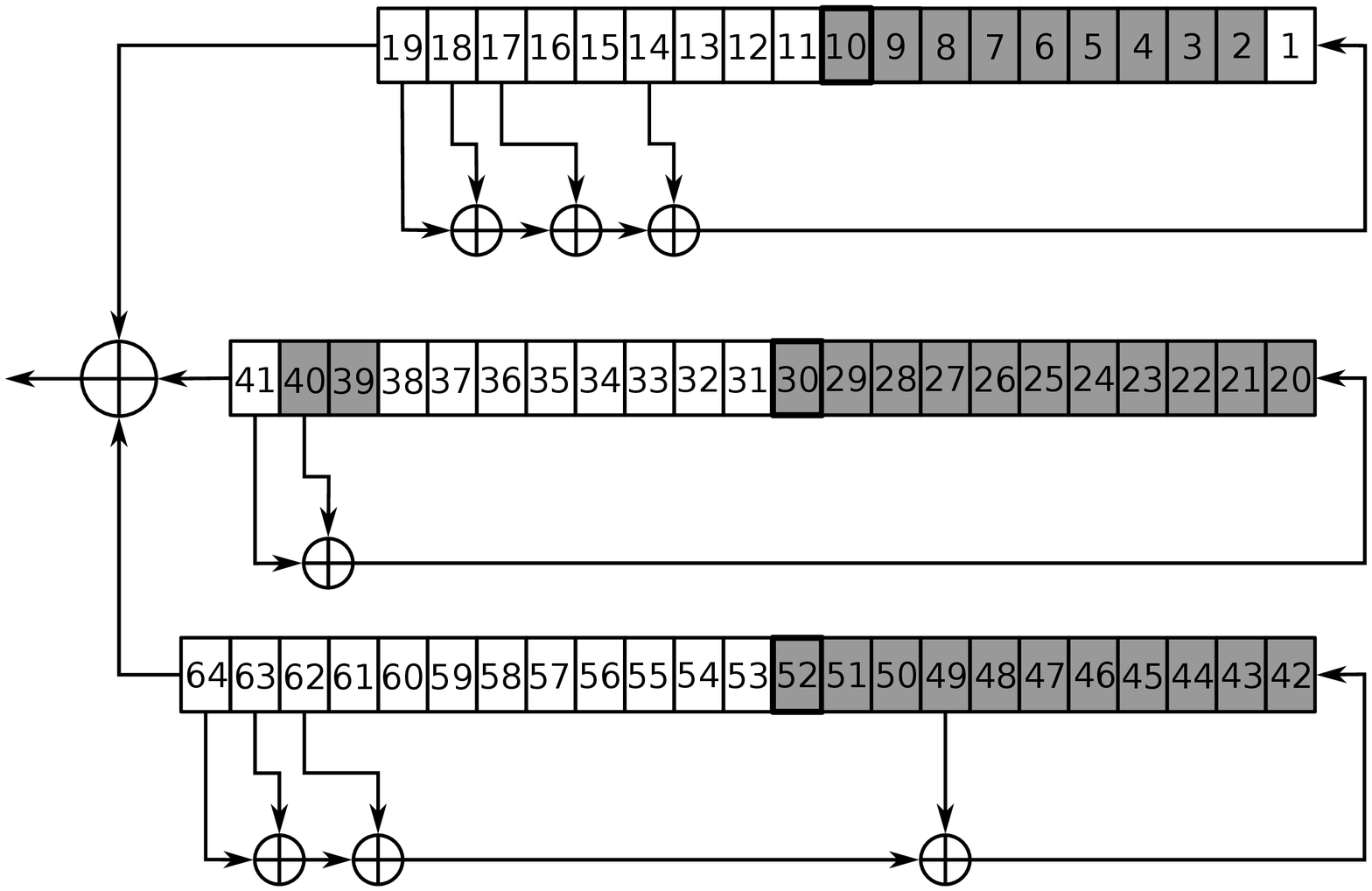}
	\caption{Decomposition set S3}
	\label{a5_1_set_S3}
\end{figure}

On Figures \ref{a5_1_set_S1}, \ref{a5_1_set_S2}, \ref{a5_1_set_S3} three decompositions sets are shown. We described the first set (further referred to as S1) in the paper \cite{DBLP:conf/pact/SemenovZBP11} earlier. This set (of 31 variables) was constructed ``manually'' based on the analysis of algorithmic features of the A5/1 generator. The second one (S2), consisting of 31 variables, was found as a result of the minimization of $F\left( \cdot \right)$ by the simulated annealing algorithm (see section 4). The third set (S3), consisting of 32 variables, was found as a result of minimization of $F\left( \cdot \right)$ by the TS algorithm. In the table \ref{a5-1_results} the values of $F\left( \cdot \right)$ (in seconds) for all three sets are shown. Note that each of sets $S_2$ and $S_3$ was found for one 114 bit fragment of keystream that was generated according to the A5/1 algorithm for a randomly chosen 64-bit secret key.

\begin{table}
\caption{Decomposition sets fot the A5/1 cryptanalysis problem and corresponding values of the predictive function. }
\label{a5-1_results}
\centering
\begin{tabular}{p{1.8cm}|p{1.8cm}|p{1.8cm}}
Set & Power of set & $F\left( \cdot \right)$\\
\hline 
\hline
S1 & 31 & 4.45140e+08 \\
\hline 
S2 & 31 & 4.78318e+08 \\
\hline 
S3 & 32 & 4.64428e+08 \\
\end{tabular}
\end{table}

\section{Cryptanalysis of the A5/1 Keystream Generator  in the Volunteer Computing Project SAT@home}

Volunteer computing \cite{DBLP:journals/jnca/DurraniS14} is a type of distributed computing which uses computational resources of PCs of private persons called volunteers. Each volunteer computing project is designed to solve one or several hard problems. When PC is connected to the project, all the calculations are performed automatically and do not inconvenience user since only idle resources of PC are used.

The first volunteer project was the GIMPS\footnote{http://www.mersenne.org/} project launched in 1996. Nowadays the most popular platform for organizing volunteer computing projects is BOINC \cite{DBLP:conf/grid/Anderson04} that was developed in Berkeley in 2002. Today there are about 70 active volunteer projects, the majority of them based on BOINC. The total performance of these projects exceeds 10 petaflops. Among the most important results obtained via volunteer computing are the discoveries of the largest prime number in the GIMPS project and of several radiopulsars in the Einstein@home\footnote{http://einstein.phys.uwm.edu/} project. 

Volunteer computing project consists of the following basic parts: server daemons, database, web site and client application. Client application should have versions for the widespread computing platforms. One of the attractive features of volunteer computing is its low cost --- to maintatin a project one needs only a dedicated server working 24/7. Another advantage of this type of computing is that volunteer project can solve some particular hard problem for months or even years with good average performance.

SAT@home\footnote{http://sat.isa.ru/pdsat/} \cite{journals/csj/Posypkin12} is volunteer computing project aimed at solving hard combinatorial problems that can be effectively reduced to SAT. It was launched on September 29, 2011 by ISDCT SB RAS and IITP RAS. SAT@home was developed with the help of BOINC platform  \cite{DBLP:conf/grid/Anderson04} and SZTAKI Desktop Grid package \cite{sztaki6003}. On February 7, 2012 SAT@home was added to the official list of BOINC projects\footnote{http://boinc.berkeley.edu/projects.php} with alpha status. Recently its status was improved to beta.

The SAT@home server uses a number of standard BOINC daemons responsible for sending and processing tasks (transitioner, feeder, scheduler, etc.). Such daemons as work generator, validator and assimilator were implemented taking into account the specificity of the project. The work generator decomposes the original SAT problem to subproblems according to the partitioning approach (see section 4). It creates 2 copies of each task in accordance with the concept of redundant calculations used in BOINC. The validator checks the correctness of the results, and the assimilator processes correct results. The client application is based on the SAT solver \textsc{MiniSat} \cite{DBLP:conf/sat/EenS03}.

The characteristics of the SAT@home project as of 12 of November 2014 are (according to BOINCstats\footnote{http://boincstats.com/}):
\begin{itemize}
	\item 3113 active PCs (active PC in volunteer computing is a PC that sent at least one result in last 30 days) about 80\% of them use Microsoft Windows OS;
	\item 1337 active users (active user is a user that has at least one active PC);
	\item	versions of the client application for CPU: Windows x86, Windows x86-64, Linux x86, Linux x86-64;
	\item average real performance: 3.2 teraflops, maximal performance: 7.9 teraflops.
\end{itemize}
The dynamics of the real performance of SAT@home can be seen at the SAT@home performance page\footnote{http://sat.isa.ru/pdsat/performance.php}.

The experiment consisting in solving 10 inversion problems of function $f_{A5/1}$ was held in SAT@home from December 2011 to May 2012. To construct the corresponding tests we used the known rainbow-tables for the A5/1 algorithm\footnote{https://opensource.srlabs.de/projects/a51-decrypt}. These tables provide about 88\% probability of success when analyzing 8 bursts of keystream (i.e. 914 bits). We randomly generated 1000 problems and applied the rainbow-tables technique to analyze 8 bursts of keystream, generated by A5/1. Among these 1000 problems the rainbow-tables could not find the secret key for 125 problems. From these 125 problems we randomly chose 10 and in the computational experiments applied the SAT approach to the analysis of first bursts of the corresponding keystream fragments (114 bits). For each SAT instance we constructed the partitioning generated by the S1 decomposition set (see Figure \ref{a5_1_set_S1}) and processed it in the SAT@home project. All 10 instances constructed this way were successfully solved in SAT@home in about 5 months (the average performance of the project at that time was about 2 teraflops). After finding the satisfying assignment the processing of the corresponding partitioning was interrupted. On average to solve one SAT instance the SAT@home project had to process about 1 billion subproblems from the corresponding partitioning.

The second experiment on the cryptanalysis of A5/1 was launched in SAT@home in May 2014. It was done with the purpose of testing the decomposition sets found automatically with the help of the procedures described in section 4. In particular we took the decomposition set S3 (see Figure \ref{a5_1_set_S3}) that was constructed for the SAT instance encoding the cryptanalysis of one particular 114-bit A5/1 keystream fragment, and then used this set to solve the series of tests with 10 similar instances. On September 26 we successfully solved in SAT@home all 10 instances from the considered series. It should be noted that in all the experiments the time required to solve the problem agrees with the predictive function value computed for the S3 decomposition set.

The problem of cryptanalysis of the generator A5/1 is also interesting because the same keystream of an arbitrary length can be generated from different secret keys. This fact was noted by Jovan Golic in \cite{Golic:1997:CAA:1754542.1754566}. We refer to these situations as ``collisions'' using the evident analogy with the corresponding notion from the theory of hash functions. The approach presented in our paper allows us to solve the problem of finding all the collisions of the generator A5/1 for a given fragment of a keystream. Using the SAT@home project collisions for several SAT instance were found (see Table \ref{Collisions}).

\begin{table}
\renewcommand{\arraystretch}{1.3}
\caption{Original keys and collisions of the generator A5/1 \mbox{(in hexadecimal format)} }
\label{Collisions}
\centering
\begin{tabular}{c|c|c}
\textbf{instance} & \textbf{original secret key} & \textbf{collision} \\ 
\hline
1 & F43FF04CD4F45660 & 7A1FF04CD4F45660 \\
\hline 
2 & B95654F2242C6DF1 & 5CAB34F2242C6DF1 \\
\hline 
3 & 67685940B034EF78 & B3B43940B034EF78 \\
\hline 
4 & 41038717BBA05E57 & FB164EC44124398E \\
\end{tabular}
\end{table}

Our computational experiments clearly demonstrate that the proposed method of automatic search for decomposition sets makes it possible to
construct SAT partitionings with the properties close to that of ``reference'' partitionings, i.e. partitionings constructed based on the analysis of algorithmic features of the considered cryptographic functions.

\section{Related Works}
In the book \cite{DBLP:series/faia/2009-185} the Boolean satisfiability problem is considered in the variety of aspects. The problems regarding the organization of the SAT solving process in distributed computing environments have been studied in \cite{DBLP:journals/jsc/ZhangBH96,DBLP:journals/pc/BlochingerSK03,DBLP:conf/sat/HyvarinenJN06,DBLP:journals/pc/ChrabakhW06,DBLP:conf/hvc/HeuleKWB11,Hyvarinen11}. The partitioning approach was analyzed in detail in \cite{DBLP:conf/sat/HyvarinenJN06,DBLP:conf/lpar/HyvarinenJN10,Hyvarinen11}. A number of approaches to estimating the runtime for Constraint Satisfaction Problem and SAT can be found in \cite{DBLP:journals/tcad/AloulSS03,DBLP:conf/aaai/KilbySTW06,DBLP:conf/sat/HaimW08,DBLP:journals/jair/XuHHL08}.

The first example of the application of the SAT approach to the cryptanalysis problems is \cite{DBLP:journals/jar/MassacciM00}. In \cite{techrep/Mcdonald07,DBLP:conf/sat/EibachPV08,DBLP:conf/sat/SoosNC09,DBLP:conf/tools/Soos10} several estimations of time required to solve SAT for CNFs encoding the Bivium cipher cryptanalysis were proposed. The ideas underlying the cited works are close to the ones we use. The main distinction between our results and that of \cite{techrep/Mcdonald07,DBLP:conf/sat/EibachPV08,DBLP:conf/sat/SoosNC09,DBLP:conf/tools/Soos10} is that the decomposition sets that were constructed in those papers were found based on either the structure of the encryption algorithm or randomly. Our main achievement consists in constructing automatic procedures to solve this problem.

The most effective in practice method of cryptanalysis of A5/1 is the Rainbow-method, partial description of which can be found on the A5/1 Cracking Project site\footnote{https://opensource.srlabs.de/projects/a51-decrypt}. In \cite{Guneysu:2008:CC:1446228.1446266} the description of a number of techniques, used in the A5/1 Cracking Project to construct Rainbow tables, was presented. The cryptanalysis of A5/1 via Rainbow tables has the success rate of approximately 88\% if one uses 8 bursts of keystream. The success rate of the Rainbow method if one has only 1 burst of keystream is about 24\%. In all our computational experiments we analyzed the keystream fragment of size 114 bits, i.e. one burst. In \cite{DBLP:conf/pact/SemenovZBP11} we described our first experience on the application of the SAT approach to A5/1 cryptanalysis in the specially constructed grid system BNB-Grid. The decomposition set used in \cite{DBLP:conf/pact/SemenovZBP11} was constructed manually based on the peculiarities of the A5/1 algorithm.
 
The grid systems aimed at solving SAT are relatively rare. In \cite{DBLP:journals/grid/SchulzB10} a desktop grid for solving SAT which used conflict clauses exchange via a peer-to-peer protocol was described. Apparently, \cite{DBLP:conf/grid/BlackB11} became the first paper about the use of a desktop grid based on the BOINC platform for solving SAT. Unfortunately, it did not evolve into a full-fledged volunteer computing project. As we noted above, the predecessor of the SAT@home was the BNB-Grid system \cite{DBLP:journals/ife/EvtushenkoPS09,DBLP:conf/pact/SemenovZBP11}, that was used to solve first large scale cryptographic problems in 2009. 

\section{Conclusions}
In the present paper we describe the volunteer computing project SAT@home designed specifically for the purpose of long-term solving of hard SAT instances. We also propose the partitioning method for SAT instances encoding inversion problems of cryptographic functions. The applicability of proposed technologies is demonstrated on the problem of cryptanalysis of the A5/1 keystream generator. In the SAT@home project we managed to solve several cryptanalysis instances for this generator. In all the experiments we analyzed only the first burst (114 bits) of keystream.

\section{Acknowledgements}
Authors thank Stepan Kochemazov for numerous valuable comments that allowed us to significantly improve the quality of the paper, Alexey Ignatiev and Irina Bogachkova for constructive feedback and helpful discussions. This work was partly supported by Russian Foundation for Basic Research (grants 14-07-00403-a and 14-07-31172-mol-a) and by the President of Russian Federation grants for young scientists (SP-1855.2012.5 and SP-3667.2013.5).

\bibliographystyle{splncs03}
\bibliography{j_network_computer_applications_fin}

\end{document}